\documentclass[prb,twocolumn,preprintnumbers,amsmath,amssymb]{revtex4}
\usepackage{graphicx}% Include figure files
\usepackage{dcolumn}% Align table columns on decimal point
\usepackage{bm}% bold math
\begin{document}
\title{Phase Behavior in Thin Films of Confined Colloid-Polymer Mixtures}
\author{Chun-lai Ren and Yu-qiang Ma$^*$}

\address{National Laboratory of Solid State Microstructures, Nanjing University, Nanjing 210093,
China\\Received   November 6, 2005; E-mail: myqiang@nju.edu.cn.}
\begin{abstract}
Using self-consistent field and density-functional theories, we
first investigate colloidal self-assembling of colloid/polymer
films confined between two soft surfaces grafted by polymers. With
the increase of colloidal concentrations, the film undergoes a
series of transitions from disordered liquid $\rightarrow$ sparse
square $\rightarrow$ hexagonal (or  mixed square-hexagonal)
$\rightarrow$ dense square $\rightarrow$ cylindric structures in
plane, which results from the competition between the entropic
elasticity of polymer brushes and the steric packing effect of
colloidal particles. A phase diagram displays the stable regions
of different in-layer ordering structures as the colloidal
concentration is varied and layering transitions as the
polymer-grafted density is decreased. Our results show a new
control mechanism to stabilize the ordering of structures within
the films.

\end{abstract}

 \maketitle
\section*{Introduction}
Dispersions of colloidal particles in polymer solutions form the
basic ingredients of a wide variety of  systems which are
essential to life and major industries.\cite{a001,a0001} In
particular, assembling colloidal nanoparticles into ordered
structures
\cite{0a,a2aa,a2bb,ad1,ad1aa,ad2,ad2aa,ad3,ad3aa,cheneeee,cheneee,chenee,chenee1,chenee2,chenee3,chenee4}
is important to the fabrication of nanomaterials with high
performance in size-dependent optical, electrical, and mechanical
properties. A major challenge in this field is to assemble
monodispersed colloids into highly ordered structures with
well-controlled sizes and shapes. Usually, the dispersed
structures depend upon the effective interaction between the
colloidal particles themselves and between the colloidal particles
and their surroundings. For example, the interaction between
colloidal particles in a solution of nonadsorbing polymers is
 influenced by the entropic excluded-volume (depletion)
forces which were first recognized theoretically by Askura and
Oosawa (AO),\cite{a4} and thus a monodispersed colloidal system
may typically crystallize into a close-packed structure or
body-centered cubic lattice with high symmetry.

Thin films made of mixtures of polymer and colloidal particles can
exhibit interesting structures, which are particularly useful for
 nano(bio)technology ranging from nanopatterning/lithography to
 changes of surface properties. Surface confinement  modifies the
  bulk behavior of the system by breaking the
symmetry of the structure, and therefore can   be used to make
low-dimensional ordering films. For  nano-sized colloidal
particles, the confinement of the suspension is a straightforward
approach for the formation of colloidal crystals. For example,
when the colloid/polymer mixtures are sandwiched between two
parallel hard surfaces,  layered structures may   form, but the
in-plane   structure  of the particles is always uniform or
close-packed. Polymer brushes grafted to surfaces can modify the
properties of the surface,\cite{a999,a999x,a11xa} which has many
 applications in colloidal stabilization  and bio-compatibility.
 Interestingly, the   grafted polymer chains may lead to
 a long-ranged repulsive force between the particles,
 which stabilizes the ordering of  the nano-sized particles
  and controls the symmetry of the structure formed  in a
desired direction. Such a control is required for a wide range of
applications using structure-dependent optical, electrical, and
mechanical properties of the colloidal crystals. It has been noted
\cite{a10,a101,a101aa,a102,a102aa} that thin films made of charged
colloidal suspensions have various layered crystalline states if
the film is confined between two parallel plates which repel the
colloidal particles. Here, we show that for colloid/polymer
mixtures, using polymer brush surfaces is an effective way of
changing  the in-plane structure of the colloidal film.

The purpose of the present article is to demonstrate a novel route
of spatially organizing the colloid dispersions in polymer
solutions confined between two polymer-grafted plates. We find
that with the deformation of grafted   chains  by
 the particles, the deformed brushes can exert an additional
interaction between nano-sized colloidal particles, which
stabilizes the formation of low symmetric ordering structures of
colloids. Such a brush-mediated colloidal effective interaction
will compete with steric (center-force) effects of colloids, which
may generate structures with more complicated lattice symmetries
and undergo the symmetry-changing transitions in plane. In the
present model, layering structures are formed due to confinement,
and more importantly, the in-layer structural transitions are
observed, depending upon the colloidal concentration and the
polymer-grafted density. The results are summarized in a phase
diagram displaying the stability regions of self-assembled
ordering phases as functions of the colloidal concentration and
the polymer-grafted density. The proposed approach is physically
achievable for the generation of novel ordered structures. A
successful example indicating the importance of such a
self-assembling mechanism is the recent experimental
verification\cite{ad3}  of  the predicated phase behavior of the
copolymer/particle mixtures in confined
geometries.\cite{add2,add3} Another motivation of the present work
is due to the fact that although many experiments have
demonstrated the success of fabricating self-assembly
nano-structures, the  understanding of the physical mechanism that
drives the self-assembly of the colloidal particles in confined
geometries is still lacking.\cite{add2,add3,add4,add5}

\section*{Model Details}

We consider a colloid suspension confined between two planar
plates separated by a distance $L_z$ along the z-axis. The two
plates which are grafted with $n_{br}$ A-type homopolymer chains,
are horizontally placed in $xy$-plane, and mechanically fixed at
$z=0$ and $z=L_z$, respectively. All B-type free homopolymer
chains in solutions are flexible with the same polymerization $N$
and statistical length $a$, and incompressible with a segment
volume $\rho _0^{-1}$ as in A-type grafted chains. The volume of
the system $V$ is $L_x\times\L_y\times L_z$, where $L_x$ and $L_y$
are the lateral lengthes  of   surfaces. The grafting density is
defined as $\sigma =n_{b}/(2\times L_x\times L_y)$; the average
volume fraction of grafted chains is $\phi_{b}=n_{b}N\rho
_0^{-1}/V$, the colloid $\phi_{c}=(1-\phi_{b}){\psi_{c}}$ where
$\psi_{c}$ is the colloidal concentration of the colloid-polymer
mixture, and the free polymer $\phi_{p}=1-\phi_{b}-\phi_{c}$.
Recently, the self-consistent field theory (SCFT) has been proven
to be powerful in calculating equilibrium morphologies in
polymeric
systems,\cite{a17,a17aa,a00,a11,a22,a33,a44,a55,a66,a13,a13aa}
while colloidal particles can be treated by density-functional
theory (DFT) \cite{a13,a13aa,a15,a16} to account for steric
packing effects of  particles. In our calculation, we use the
hybrid SCFT/DFT approach developed in refs 41 and 42, and the free
energy F for the present system is given by
\begin{eqnarray}
\frac{NF}{\rho _0 k_B T V}&=&-\phi_{b}\ln (\frac{Q_{b}}{V\phi_{b}})-\phi_{p}\ln (\frac{Q_{p}}{V%
\phi_{p}})-\frac{\phi_{c}}{\alpha}\ln (\frac{Q_{c}\alpha}{V\phi_{c}})   \nonumber\\
 &&+\frac 1V\int
d\mathbf{r}[\chi _{bp}N\varphi _{b}({\bf r})\varphi _p({\bf r})%
+\chi _{bc}N\varphi _{b}({\bf r})\varphi_{c}({\bf r})\nonumber\\
&&+\chi _{pc}N\varphi _{p}({\bf r})\varphi_{c}({\bf r})
 -W_{b}({\bf r})\varphi _{b}({\bf r})- W_{p}({\bf r})\varphi _{p}({\bf r})\nonumber\\
 &&- W_{c}({\bf r})\rho_{c}({\bf r})
 -\xi({\bf r}) (1-\varphi _{b}({\bf r})- \varphi _{p}({\bf
r})-\varphi _{c}({\bf r}))\nonumber\\
 && +\rho_{c}\Psi_{hs}(\overline{\varphi
}_{c})]\;,
\end{eqnarray}
where $k_B$ is the Boltzmann constant, and  $T$ is the
temperature. $\alpha$ is the volume ratio of the colloidal
particle of radius $R$ to polymer chain: $\alpha = \frac{ v_R
\rho_{0}}{N}$, where $v_R=\frac{4}{3}\pi R^3$. $\varphi _{b}(\bf
r)$, $\varphi _{p} (\bf r)$, and $\varphi _{c}(\bf r)$ are  the
local volume fractions  of brushes, free polymers,  and colloids,
respectively. $\xi(\bf r) $ is the potential field that ensures
the incompressibility of the system for dense colloid-polymer
mixtures. $\rho_c$ stands for the particle center distribution,
and the local particle volume fraction is then given by
$\varphi_{c}(\mathbf{r})=\frac{\alpha}{v_{R}} \int_{|{\bf
r'}|<R}d\mathbf{r'}\rho_{c}(\mathbf{r}+\mathbf{r'})$.
\cite{a13,a13aa}  The $\chi$'s are the Flory interaction
parameters between the different chemical species.
$Q_{b}=\int d\mathbf{r}q_1\left( \mathbf{r,}s\right) q_1^{+}\left( \mathbf{r,%
}s\right)$ and $Q_{p}=\int d\mathbf{r}q_2\left(%
\mathbf{r,}1\right)$ are single-chain partition functions  for
brushes and free polymers under the self-consistent fields
$W_{b}(\bf r)$ and $W_{p}(\bf r)$, respectively. $Q_{c}=\int
d\mathbf{r}\exp[-W_{c}(\mathbf{r})]$ is the partition function for
colloids under the self-consistent field $W_{c}(\bf r)$. The
end-segment distribution functions $q_{i}(\mathbf{r},s)$ and
$q_{i}^{+}(\mathbf{r},s)$ represent the probability of finding
monomer s at position $\bold r$ respectively from two distinct
ends of chains, which satisfy modified diffusion equations $
\frac{\partial q_{i}}{\partial s}=\frac{a^2 N}6\nabla
^2q_{i}-W_{i}({\bf r}) q_{i}$ , and $ \frac{\partial
q_{i}^{+}}{\partial s}=-\frac{a^2 N}6\nabla ^2q_{i}^{+}+W_{i}({\bf
r}) q_{i}^{+} $ . For the grafted chains in the field $W_{b}(\bf
r)$, the initial condition is $ q_1(x,y,z=0$ $or$ $L_z,0)=1$,
$q_1(x,y,z\neq 0$ $or$ $L_z,0)=0$, and $q_1^{+}(x,y,z,1)=1,$ which
means that the end  of brush chains can move on the plates,
although the total number of chains on surfaces is fixed. These
have been called liquid brushes, in contrast to solid brushes
where the immobile chains are anchored onto the
surfaces.\cite{a14}  Here we consider the liquid brush case
because of its wide-ranging applications in colloidal and
biological systems. For free polymers, the initial condition is
$q_2(x,y,z,0)=1$. The last term in Eq.(1) is DFT term \cite{a15}
accounting for the steric interaction between particles, and the
excess free energy $\Psi(\overline{\varphi}_c)$ per particle is
from the Carnahan-Starling function \cite{a16} $
\Psi_{hs}(x)=\frac{4x-3x^{2}}{(1-x)^{2}}\;,$ with the weighted
particle density  $ \overline{\varphi}_c({\bf r})$,\cite{a13}
$\overline{\varphi}_{c}(\mathbf{r})=\frac{\alpha}{v_{2R}}\int_{|{\bf
r'}|<2R}d\mathbf{r'}\rho_{c}(\mathbf{r}+\mathbf{r'})$, where
$v_{2R}$ is the volume of a sphere of radius $2R$.

By minimizing the free energy in Eq. (1) with respect to
$W_{b}(\bf r)$, $W_{p}(\bf r)$, $W_{c}(\bf r)$, $\varphi _{b}(\bf%
r)$, $\varphi _{p}(\bf r)$, $\varphi _{c}(\bf r)$, and $\xi(\bf%
r)$, we obtain a set of self-consistent equations describing the
equilibrium morphology of films, which can numerically be solved
by  the real space combinatorial screening algorithm of Drolet and
Fredrickson.\cite{a17,a17aa}  The free energy minimization of the
system with respect to the selected simulation sizes is
performed.\cite{a18}   We fix $N=100$, $L_z=50a$, and the colloid
diameter $2R=0.6R_{0}$, where $R_{0}\equiv aN^{\frac{1}{2}}$
characterizes the natural size of free polymers (typically $10
\sim 100$ nm). On the other hand, based on the assumption that the
interaction between different chemical species should be small
enough  and polymer brushes are chemically neutral to free
polymers and colloids, we choose  the Flory-Huggins interaction
parameters $\chi_{pc}N=5.0 $  and $\chi_{bp}N=\chi_{bc}N=8.0$. All
the sizes are in units of $a$.
\begin{figure*}
\includegraphics[width=17.5cm]{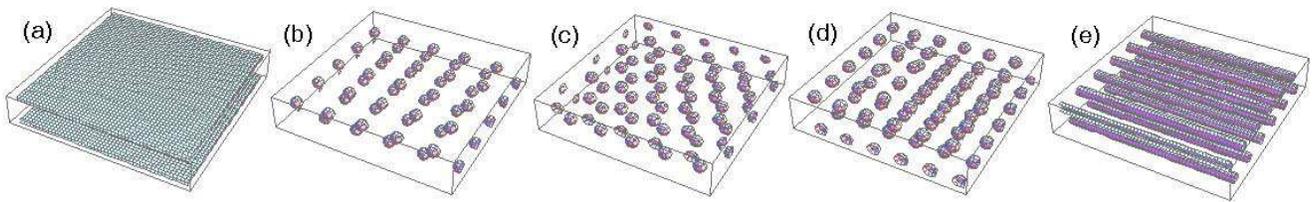}
\caption{Distributions of colloidal particles  with increasing the
colloidal concentration for the brush density $\sigma=0.2$ and
$L_x=L_y=50$. The colloidal particles between two layers are
alternatively packed each other. (a) $\psi_{c}=0.29$, (b)
$\psi_{c}=0.30$, (c) $\psi_{c}=0.35$, (d) $\psi_{c}=0.38$, and (e)
$\psi_{c}=0.45$.}
\end{figure*}
\begin{figure*}
\includegraphics[width=17.5cm]{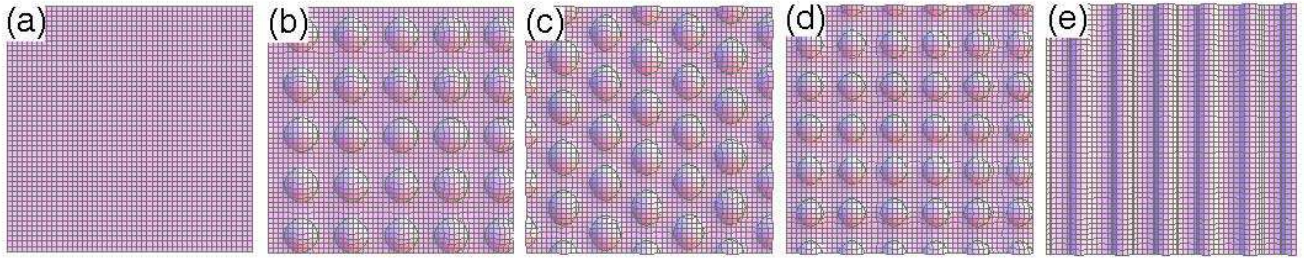}
\caption{  Top views of the density profiles of polymer brushes.
Parameters are the same as in Figure 1.}
\end{figure*}

\section*{Results and Discussion}

We first explore the formation of ordered structures of colloidal
crystallization in confined colloidal solution film.  Figure 1
shows a series of particle-dispersed morphologies of varying the
colloidal concentration  $\psi_{c}$, and Figure 2 is the top views
of profiles for polymer brushes grafted to one of the plates. We
observe the formation of two layers of colloidal particles which
are close to the polymer brush. When the colloidal concentration
is small (for example, $\psi_{c}=0.29$, as in Figure 1(a)),
colloids distribute uniformly in plane, and the brush-formed
interfaces remain level(Figure 2(a)). With increasing the
colloidal concentration, colloidal crystallization may appear, and
inhomogeneous colloidal distribution will lead to the deformation
of polymer brushes: caving where colloids exist and protruding
where not. The deformed brushes prefer to stabilize the formation
of low symmetric ordering structures of colloids, and with
increasing the colloidal concentration, the particles grow
anisotropically along one-dimensional direction.\cite{a19}   We
see from Figures 1(b)-1(e) that for different colloidal
concentrations, colloids can assemble into sparse square,
hexagonal, dense square, and cylindric structures, depending on
the competition between the entropic effect of polymer brushes and
steric packing effect of the particles. Figures 2(b)-2(e) show the
corresponding shapes of deformed brush  surface due to respective
colloidal structures. When $\psi_{c}=0.30$, the average distance
between colloidal particles is relatively large, and thus the
brush-mediated interaction between colloids dominates over the
steric packing effect  of colloids, leading to in-layer square
structures for possibly large conformational entropy of polymer
brushes(Figure 2(b)). When $\psi_{c}=0.35$, the stable state
becomes hexagonal one(Figure 1(c) and   2(c)), indicating that
colloidal steric packing effects dominate. However, a dense square
structure appears at $\psi_{c}=0.38$ (Figure 1(d)), due to the
requirement of entropic elastic release of deformed brushes.
Further increase of colloids greatly limits the conformational
space of brushes, and the entropy of brushes may be released by
the formation of cylinder arrangement of colloids, because in this
case, the free ends of grafted chains can at least move freely
between parallel cylinders, as shown in Figure 1(e) and
   2(e) when $\psi_{c}=0.45$.
Figure 2 shows that the  polymer brush is soft, and its density
variation changes with the colloidal concentration. The range of
the density variation in the direction perpendicular to the
substrate is comparable to the size of the colloidal particles.
The brush density close to the substrate remains unchanged with
varying colloidal concentrations for a given grafting density.

\begin{figure}
\includegraphics[width=9cm]{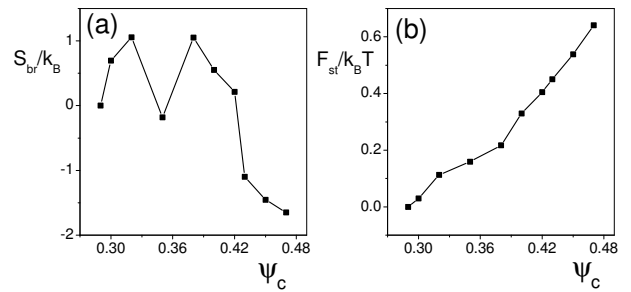}
\caption{ (a) A grafted-chain entropy ($\frac{S_{br}}{k_B}$) vs
the colloidal concentration; (b) the averaged steric energy per
colloidal particle ($\frac{F_{st}}{k_BT}$)
 vs the colloidal concentration.}
\end{figure}

Figure 3(a) shows the entropy of a grafted chain as a function of
the colloidal concentration $\psi_{c}$. We set the grafted chain's
entropy $\frac{S_{br}}{k_B}=0$, when the dispersion of colloidal
particles is uniform in plane ($\psi_{c}=0.29$). The entropy of
polymer brushes, with colloidal concentrations varied from
$\psi_{c}=0.29$ to $\psi_{c}=0.32$, increases with the formation
of colloidal square structures. However, the brush entropy curve
suddenly decreases with the influence of steric interaction of
colloids, and a hexagonal structure is formed when
$\psi_{c}=0.35$. With further increase of colloidal
concentrations, the deformed brushes begin to dominate over the
steric energy of colloids again, which prefers to release the
conformational space of brushes with the gradual formation of
square structures. The brush entropy arrives at the maximum at
$\psi_{c}=0.38$, corresponding to the formation of dense square
lattice structure (Figure 1(d)). After that, brush entropy
decreases with the increase of colloidal concentrations, and
colloidal cylinder structures begin to form. The complete
cylindric structure appears at $\psi_{c}=0.42$, and remains
unchanged even for relatively larger colloidal concentrations.
Usually, the entropic gain or loss due to the conformational
change of polymer chains may signify the structural transitions of
the polymeric system. For example, Balazs and
co-workers\cite{a13,a13aa}  studied an AB diblock-nanoparticle
composite, in which the particles and diblock chains form
spatially periodic structures. They observed a similar behavior of
A-block entropic free energy when the volume fraction of the
particles is increased.\cite{a13}   In their work, however, the
conformational entropy of the A-block only shows the stretching
and compressing of the A block chains along the chain direction,
because they considered the formation of spatially periodic
structures in two-dimensional copolymer-nanoparticle composites in
the plane of the polymer chains. On the contrary, in the present
three-dimensional confined mixtures of colloidal particles and
free homo-polymer chains, we emphasis a control mechanism in the
lateral direction of the colloidal suspension driven by the
deformed chains in the polymer brush.  Actually, the entropic free
energy in Figure 3(a) depicts the shape changes of brush surface
(Figure 2) during the variation of the colloidal concentration for
\begin{figure*}
\includegraphics[width=17.5cm]{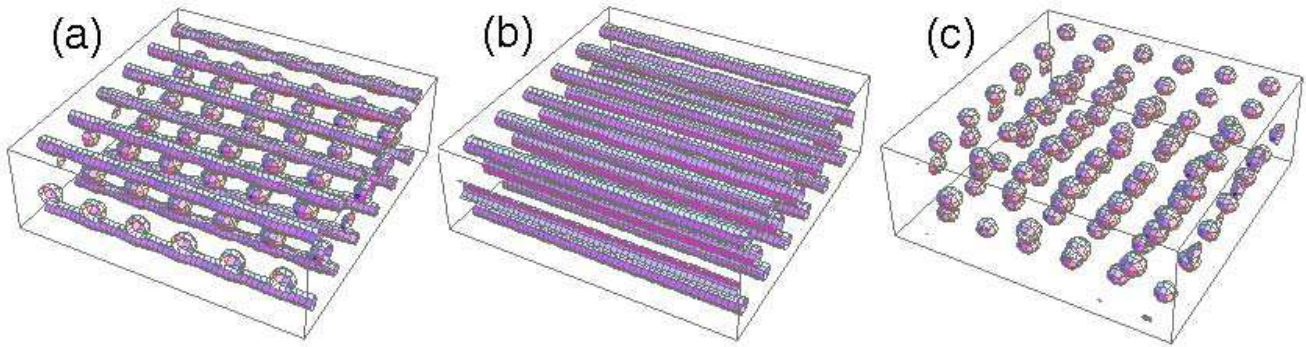}
\caption{   The formation of three-layer structures: (a)
sphere-cylinder mixture when $\sigma=0.17$ and $\psi_{c}=0.36$;
(b) cylinder structure when $\sigma=0.17$ and $\psi_{c}=0.42$; (c)
spherical structure when $\sigma=0.16$ and $\psi_{c}=0.32$.}
\end{figure*}
\begin{figure}
\includegraphics[width=9cm]{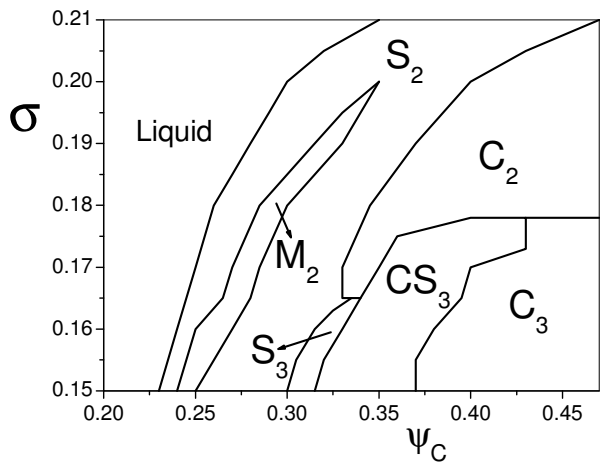}
\caption{The phase diagram as a function of the grafting density
and the colloidal concentration.  $S_{2}$: two-layered sparse or
dense square structure; $M_{2}$: two-layered hexagonal or
 mixed square-hexagonal structure; $C_{2}$: two-layered   cylindric
structure; $S_{3}$: three-layered spherical structure; $CS_{3}$:
three-layered   cylindric-spherical mixed structure; $C_{3}$:
three-layered  cylindric structure.}
\end{figure}
a given grafting density $\sigma=0.2$. The configurational entropy
contribution of the grafted chains originates from the free ends
of the brushes comparable in size to the colloidal particles,
since the bottom part of the chains close to the substrate is
strongly stretched and remains nearly unchanged with varying
colloidal concentration for a fixed grafting density. This means
that the entropic effect of the deformed brushes by the particles
comes mainly from the lateral conformation fluctuations of the
free chain ends, and that the averaged conformational entropy
contribution of the polymer brushes along the chain-stretching
direction can be ignored.  Such an entropic effect competes with
the steric packing effect of the particles in the plane, leading
to structure changes in the plane of the colloidal particles.
Here, a significant advantage of using soft polymer-grafted
substrate is that the deformed brush provides an in-plane
long-ranged anisotropic interaction between the colloidal
particles. In Figure 3(b), the average steric energy per colloidal
particle varies with the increase of colloidal concentration,
which clearly reflects the excluded-volume effect of colloids. As
a whole, steric energy increases with increasing the colloidal
concentration $\psi_{c}$. Interestingly, we find that the slope of
the curve has apparent difference in different ranges of
$\psi_{c}$. Particularly in the region $\psi_{c} \sim 0.32-0.38$
where the morphology inclines to adopt hexagonal or mixed
square-hexagonal structure, the slope is distinctly lower than
that of other regions. This means that the steric energy can be
released by forming hexagonal or mixed hexagonal/square
structures.

Further, we also observe the appearance of layering transitions
with the increase of effective film thickness via decreasing the
height of polymer brushes.
 The effective
thickness of film can be defined by subtracting the brush height
from the distance $L_z$ between two substrates, i.e.,  $d_{eff} =
L_{z}-2h$, where the dry brush height  $h$ satisfies $h=\sigma
N/\rho_{0}$ in the incompressible system of brush and homopolymer
molecules.\cite{add1}   Thus,  the effective thickness $d_{eff}$
of the film is changeable, and increases with decreasing the
polymer-grafted density $\sigma$. Figures 4(a)-4(c) show the
possible three-layer structures such as mixed sphere-cylinder,
complete cylinder, and even sphere structures for the selected
values of $\sigma$ and $\psi_c$. To clarify the phase stability of
different ordering structures,\cite{a188}  we calculate the phase
diagram as functions of the grafting density and the colloidal
concentration, as shown in Figure 5. Here, we choose the grafting
density $\sigma$ within the range [0.15, 0.21] for two reasons: on
the one hand, the entropic change of deformed  brushes should be
large enough to cause effective brush-mediated interaction between
colloids, requiring that the value of $\sigma$ can not be too
small; on the other hand, the effective film thickness should be
thick enough (namely $\sigma$ is reasonably small) to ensure the
possibility of colloidal crystallization. We see from Figure 5
that when the grafting density  is relatively large, the
two-layered colloidal structure undergoes disordered liquid,
square, and cylinder ones. In this case, the brush-mediated
interaction always dominates, which drives the self-assembly of
colloids  into square lattice at small $\psi_c$ and cylinder
structure at larger $\psi_c$. The hexagonal phase disappears at
the middle concentrations of colloids. For intermediate values of
polymer-grafted densities ($\sigma=0.21 \sim 0.18$), two-layered
structure remains unchanged, but a reentrant structure transition
between sparse square $\rightarrow$ hexagonal (or
square-hexagonal) $\rightarrow$ dense square lattice structures
appears as a result of  the competition between brush-mediated
interaction and steric packing effects of colloids. When $\sigma$
is less than $0.18$, layering transition may appear for
sufficiently large values of $\psi_{c}$. The three-layered
sphere-cylinder structure ($CS_3$ phase) first appears, where
cylinder structures closing to  brush surfaces  are formed because
of colloidal particles favoring the aggregation onto brush
surfaces at first, and the middle layer still retains discrete
spherical structure. With increasing the colloidal concentration,
 the  cylinder-forming  monolayer becomes a template which drives
the ordering of other colloidal particles in the middle of the
film with the help of the deformation of free polymers in the
suspension, and then the three-layered cylinder structure ($C_3$)
appears. If $\sigma$ is decreased below $0.165$, three-layered
spherical structure ($S_3$ phase) may be formed for intermediate
ranges of $\psi_c$, before the appearance of mixed sphere-cylinder
structures. The  phase diagram shown in Fig. 5 also provides
useful information about other structural changes of the system.
For example, one can observe the structural and layering
transitions with  increasing effective film thickness
(equivalently  expanding the distance between the two plates) via
decreasing grafting density $\sigma$ for fixed colloidal
concentration $\psi_c$. If one changes the distance between the
two plates while  increasing the colloidal concentration, the
colloidal structural transitions can be obtained along a tilted
direction from the left to right in Figure 5.

\section*{Conclusions}

We study the phase behavior in thin films of confined
colloid/polymer mixtures. Depending on the competition between
brush-mediated interaction and steric packing effects of colloids,
the colloidal self-assembly can experience a series of in-layer
symmetry-changing transitions between square lattice, hexagonal,
and cylinder structures, which can be induced by changing the
colloidal concentration. Furthermore, the layering transitions can
be obtained by changing the film thickness of the system. We find
that the colloidal particles lying on the polymer brush form a
monolayer, in which the deformed chains produce an in-plane
long-ranged anisotropic interaction, which controls the ordering
of the colloidal particles.  This layer of ordered particles then
becomes a template on which the other colloidal particles in the
middle of the film grow into a large ordered structure with the
help of free polymers in the region.  The results may provide a
helpful guide for fabricating the functionally useful
microstructures in materials science of complex liquids. The
suggested approach is also suitable to understanding colloidal
self-assembling on the other soft walls such as the packing of
biomacromolecules confined within the bio-membranes.

\section*{Acknowledgment} This work was supported by the National Natural
Science Foundation of China, No. 10334020, No. 10021001, No.
20490220, and No. 10574061.


\begin{references}
\bibitem{a188}To search the stable phases of the film, we first ascertain the
symmetries of possible  structure formation, and then calculate
the free energy of the film. Furthermore, the total free energy
minimization of the system with respect to the selected simulation
sizes is required by adjusting the lateral dimension \cite{a18}.
\end{references}
\end{document}